\documentclass[seceqn]{elsart}

\usepackage{graphicx}

\usepackage{amssymb}
\usepackage{mathrsfs}

\usepackage{amsmath}

\DeclareMathOperator{\tr}{tr}

\DeclareMathOperator{\Real}{Re}
\DeclareMathOperator{\Imag}{Im}

\DeclareMathOperator{\diag}{diag}

\begin{document}

\begin{frontmatter}



\title{Lattice formulation of 2D $\mathcal{N}=(2,2)$ SQCD based on the B model
twist}


\author{Daisuke Kadoh}
\ead{kadoh@riken.jp}
\address{Theoretical Physics Laboratory, RIKEN, Wako 2-1, Saitama 351-0198,
Japan}
\author{Fumihiko Sugino}
\ead{fumihiko\_sugino@pref.okayama.lg.jp}
\address{Okayama Institute for Quantum Physics, Kyoyama 1-9-1,%
Okayama 700-0015, Japan}
\author{Hiroshi Suzuki}
\ead{hsuzuki@riken.jp}
\address{Theoretical Physics Laboratory, RIKEN, Wako 2-1, Saitama 351-0198,
Japan}

\begin{abstract}
We present a simple lattice formulation of two-dimensional $\mathcal{N}=(2,2)$
$U(k)$ supersymmetric QCD (SQCD) with $N$~matter multiplets in the fundamental
representation. The construction uses compact gauge link variables and
exactly preserves one linear combination of supercharges on the two-dimensional
regular lattice. Artificial saddle points in the weak coupling limit and the
species doubling are evaded without imposing the admissibility. A perturbative
power-counting argument indicates that the target supersymmetric theory is
realized in the continuum limit without any fine tuning.
\end{abstract}

\begin{keyword}
Supersymmetry \sep lattice gauge theory
\PACS 11.15.Ha \sep 11.30.Pb \sep 11.10.Kk
\end{keyword}
\end{frontmatter}

\section{Introduction}
\label{sec:1}
After seminal work by Kaplan et.\
al.~\cite{Kaplan:2002wv,Cohen:2003xe,Cohen:2003qw}, there have recently
appeared various lattice formulations of extended supersymmetric gauge
theories~\cite{Sugino:2003yb,Sugino:2004qd,D'Adda:2004jb,Sugino:2004uv,
Catterall:2004np,Catterall:2005fd,Kaplan:2005ta,D'Adda:2005zk,Sugino:2006uf,
Endres:2006ic,Giedt:2006dd,D'Adda:2007ax,Catterall:2007kn,Matsuura:2008cfa,
Sugino:2008yp,Unsal:2008kx,Kikukawa:2008xw}.\footnote{See
Refs.~\cite{Unsal:2006qp,Damgaard:2007be,Takimi:2007nn,Damgaard:2007xi,
Damgaard:2007eh,Damgaard:2008pa} for relationships among these formulations
and Refs.~\cite{Giedt:2003ve,Giedt:2003vy,Giedt:2004tn,Onogi:2005cz} for
related study. See~Ref.~\cite{Giedt:2007hz} for review.}
A common feature of these lattice formulations is that at least one fermionic
symmetry~$Q$, that is a linear combination of supersymmetry charges, is
manifestly preserved even with finite lattice spacings. This could be possible,
if $Q$ is nilpotent and the continuum action $S$ is $Q$-exact, $S=QX$. Such $Q$
is thus naturally identified with the BRST supercharge in topological field
theory~\cite{Witten:1988ze,Witten:1990bs}. In fact, formulations
in~Refs.~\cite{Sugino:2003yb,Sugino:2004qd,D'Adda:2004jb,Sugino:2004uv,
Catterall:2004np,Catterall:2005fd,D'Adda:2005zk,Sugino:2006uf,D'Adda:2007ax,
Catterall:2007kn,Sugino:2008yp,Unsal:2008kx,Kikukawa:2008xw} start with a
topological field theoretical representation of a target continuum theory.
In lower-dimensional systems in particular, because of this exact fermionic
symmetry~$Q$, one expects that the full supersymmetry is restored in the
continuum limit without (or with a little) fine tuning. Quite recently, this
expectation on supersymmetry restoration was clearly
confirmed~\cite{Kanamori:2008bk} (by means of a Monte Carlo simulation) in a
lattice formulation of two-dimensional (2D) $\mathcal{N}=(2,2)$ supersymmetric
Yang-Mills theory (SYM) of~Ref.~\cite{Sugino:2004qd}.

From an extended supersymmetric theory, one can construct a topological field
theory by a procedure called \emph{topological twist}, that is to define
a new rotational group (the twisted rotation) by a particular combination of
the original spacetime rotation and an internal $R$-symmetry. The above BRST
charge~$Q$ transforms as a scalar under the twisted rotation. However, if one
does not regard the twisted rotation as a real spacetime rotation, as the
standpoint we take here, the topological twist is nothing but simple relabeling
of dynamical variables in the original supersymmetric action on a flat
spacetime; it thus does not change the physical content of the theory. This
procedure is nevertheless useful to find the above $Q$ transformation and a
$Q$-exact form of the action in the continuum theory.

In this paper, we present a lattice formulation of 2D $\mathcal{N}=(2,2)$
$U(k)$ supersymmetric QCD (SQCD) with $N$ matter multiplets in the fundamental
representation. For 2D $\mathcal{N}=(2,2)$ theories, there are two possible
ways of topological twist. One is the so-called A~model twist, with which the
twisted rotation is defined as a diagonal $U(1)$ subgroup of the product of the
original 2D rotation, $SO(2)\simeq U(1)$, and the internal $U(1)_V$ symmetry.
Another is the B~model twist, with which one takes the diagonal $U(1)$ part of
the product of the 2D rotation $U(1)$ and the internal $U(1)_A$ symmetry. See,
for example, Ref.~\cite{Witten:1993yc}. For a different twist, a different
combination of supercharges is regarded as the BRST charge~$Q$. With our
present standpoint, as noted above, these two twists are just different
relabeling of dynamical variables and thus they are completely equivalent in
continuum theory.\footnote{In the context of the topological field theory, one
does not consider the B~model twist of the 2D $\mathcal{N}=(2,2)$ $U(k)$ SQCD
with $N$ fundamental matter multiplets, because the $U(1)_A$ symmetry is
anomalous in this system and consequently the twisted rotation becomes
anomalous. The B~model twist of this system is completely legitimate in our
present context because we do not regard the twisted rotation as a real
spacetime rotation.} However, resulting lattice theories can differ and they
have their own drawback and advantage. In this classification, lattice
formulations of~Refs.~\cite{Kaplan:2002wv,Cohen:2003xe,D'Adda:2004jb,
D'Adda:2005zk,Endres:2006ic,Matsuura:2008cfa} can be regarded as those based on
the B~model twist (see also Ref.~\cite{Catterall:2008dv}), while formulations
of~Refs.~\cite{Sugino:2003yb,Sugino:2004qd,Sugino:2004uv,Catterall:2004np,
Sugino:2006uf,Sugino:2008yp,Kikukawa:2008xw} are regarded as those from the
A~model twist.\footnote{See~Refs.~\cite{Catterall:2007kn,Unsal:2006qp} for
related issues.} This classification in terms of the topological twist is
sometimes very useful. For example, this explains why it was rather nontrivial
to incorporate the superpotential in lattice formulations
of~Refs.~\cite{Endres:2006ic,Matsuura:2008cfa}; they correspond to the B~model
twist and, with the B~model twist, the holomorphic part of the superpotential
cannot be written as a $Q$-exact form, although it is $Q$-closed.

Here, our intention is to provide a \emph{simple\/} lattice formulation of 2D
$\mathcal{N}=(2,2)$ $U(k)$ SQCD with $N$ fundamental matter multiplets (and
no anti-fundamental multiplet) by using compact gauge link variables on the 2D
regular lattice. For this, we adopt the \emph{B~model twist\/} picture; the
point is the $Q$-transformation law in the matter sector.

Lattice formulations of 2D $\mathcal{N}=(2,2)$ $U(k)$ SQCD
in~Refs.~\cite{Sugino:2008yp,Kikukawa:2008xw} (those also use compact gauge
link variables) are based on the A~model twist. With the A~model twist, the
superpotential is $Q$-exact and it is almost straightforward to incorporate the
superpotential in the formulation; this is an advantage of the A~model twist
picture. However, as encountered in these references, it is rather tricky to
avoid the species doubling in the matter sector with the A model twist. With
the A model twist, the $Q$ transformation of chiral fields in the fundamental
representation, for example, has a rather symmetrical structure as the $Q$
transformation of anti-chiral fields in the \emph{anti-fundamental\/}
representation. See~Eq.~(2.17) of~Ref.~\cite{Sugino:2008yp}. For this reason,
the Wilson term introduced to lift the species doublers inevitably mixes the
fundamental fermions (their number is~$n_+$) and the anti-fundamental fermions
(their number is~$n_-$) as long as the Wilson term is compatible with the
$Q$-symmetry. This forced one to take $n_+=n_-$ in~Ref.~\cite{Sugino:2008yp} in
which the conventional Wilson Dirac operator was used.
In~Ref.~\cite{Kikukawa:2008xw}, to remove the restriction~$n_+=n_-$, the
overlap Dirac operator~\cite{Neuberger:1997fp,Neuberger:1998wv} was utilized on
the analogy of a complete chirality
separation~\cite{Luscher:1998pqa,Niedermayer:1998bi} based on the
Ginsparg-Wilson relation~\cite{Ginsparg:1981bj,Hasenfratz:1997ft}. The
construction in Ref.~\cite{Kikukawa:2008xw} is so far a unique lattice
formulation of 2D $\mathcal{N}=(2,2)$ SQCD with $n_+\neq n_-$ that can
incorporate the superpotential.

The use of the overlap Dirac operator, however, requires the admissibility
condition~\cite{Hernandez:1998et,Luscher:1998du,Neuberger:1999pz} on gauge link
variables and imposition of the admissibility complicates the lattice action.
Of course, the implementation of the overlap Dirac operator itself (especially
for odd $n_+$ or~$n_-$) is practically cumbersome.

On the other hand, with the B~model twist, left- and right-handed components of
the fundamental fermions are transformed in a symmetrical way under~$Q$;
see~Eq.~(\ref{twoxseven}) below. It is thus expected that we can separately
treat the fundamental fermions from the anti-fundamental fermions while
keeping $Q$-invariance. This is the basic idea of this paper which allows a
simple lattice action.

However, with the B~model twist, the superpotential cannot be expressed as a
$Q$-exact form and, quite unfortunately, we could not find a lattice
discretization of the superpotential term that is invariant under our lattice
$Q$~transformation. This restricts the applicability of our lattice
formulation. Nevertheless, even without the superpotential, it is discussed that
2D $\mathcal{N}=(2,2)$ $U(k)$ SQCD with $N$ fundamental matter multiplets
possesses rich physical contents. The low energy effective theory would be
given by the Grassmannian $G(k,N)$ supersymmetric nonlinear sigma
model~\cite{Witten:1993yc,Witten:1993xi}, in which one expects the spontaneous
chiral symmetry breaking $\mathbb{Z}_{2N}\to\mathbb{Z}_2$ (the $U(1)_A$ symmetry
is anomalous to be broken to $\mathbb{Z}_{2N}$) and a dynamical generation of a
mass gap~\cite{D'Adda:1978uc,Witten:1978bc,D'Adda:1978kp,Abdalla:1984en,%
Cecotti:1992rm}. It would be quite interesting to investigate these quantum
phenomena by using Monte Carlo simulations on the basis of the present lattice
formulation.

\section{The continuum target theory in the B model twist picture}
We extensively follow the notational convention of Ref.~\cite{Sugino:2008yp}
for 2D $\mathcal{N}=(2,2)$ SQCD.\footnote{In particular, we assume that the
generators of $U(k)$ are hermitian and normalized as
$\tr(T^{\mathtt{A}}T^{\mathtt{B}})=(1/2)\delta^{\mathtt{AB}}$, where
$\mathtt{A}$ and~$\mathtt{B}$ run from 0 to~$k^2-1$. The color components of
adjoint fields are defined by $(\text{field})(x)=\sum_{\mathtt{A}}
(\text{field})^{\mathtt{A}}(x)T^{\mathtt{A}}$. In the continuum theory, $D_\mu$
denote the covariant derivatives with respect to the gauge potentials~$A_\mu$;
$D_\mu=\partial_\mu+i[A_\mu,\cdot]$ for adjoint fields and
$D_\mu\Phi_{+I}=\partial_\mu\Phi_{+I}+iA_\mu\Phi_{+I}$
and~$D_\mu\Phi_{+I}^\dagger=\partial_\mu\Phi_{+I}^\dagger
-i\Phi_{+I}^\dagger A_\mu$ for a field~$\Phi_{+I}$ in the fundamental
representation.}
From the $(U(1),U(1)_A)$ charges of spinorial fields, where the first $U(1)$ is
the spacetime rotation, it turns out that
\begin{equation}
      Q'\equiv-\frac{1}{\sqrt{2}}(\bar Q_L+\bar Q_R)
\label{twoxone}
\end{equation}
can be taken as a BRST supercharge with the B~model twist. We also use
different linear combinations of variables in the $\mathcal{N}=(2,2)$ gauge
multiplet from~Ref.~\cite{Sugino:2008yp}:
\begin{align}
   &\psi_0'\equiv\frac{1}{\sqrt{2}}(\lambda_L-\lambda_R),\qquad
   \psi_1'\equiv\frac{i}{\sqrt{2}}(\lambda_L+\lambda_R),
\nonumber\\
   &\chi'\equiv\frac{1}{\sqrt{2}}(\bar\lambda_L+\bar\lambda_R),\qquad
   \eta'\equiv\frac{i}{\sqrt{2}}(\bar\lambda_L-\bar\lambda_R),
\nonumber\\
   &X_0\equiv-\frac{1}{2}(\phi+\bar\phi),\qquad
   X_1\equiv-\frac{i}{2}(\phi-\bar\phi),
\nonumber\\
   &D'\equiv D-D_\mu X_\mu,
\nonumber\\
   &\mathcal{A}_\mu\equiv A_\mu-iX_\mu,\qquad
   \mathcal{A}_\mu^\dagger=A_\mu+iX_\mu,
\label{twoxtwo}
\end{align}
where $D_\mu X_\mu=\partial_\mu X_\mu+i[A_\mu,X_\mu]$.
Compare these with Eq.~(2.4) of Ref.~\cite{Sugino:2008yp}. With the B~model
twist, complexified gauge potentials in the last line naturally appear as we
will see below. We thus introduce the covariant derivatives with respect to the
complexified gauge potential~$\mathcal{A}_\mu$. For the adjoint representation,
it is defined by
\begin{equation}
   \mathcal{D}_\mu\equiv\partial_\mu+i[\mathcal{A}_\mu,\cdot].
\end{equation}
For a generic field in the fundamental representation of $U(k)$, $\Phi_{+I}$
($I=1$, \dots, $N$),
\begin{align}
   &\mathcal{D}_\mu\Phi_{+I}
   \equiv\partial_\mu\Phi_{+I}+i\mathcal{A}_\mu\Phi_{+I},
\nonumber\\
   &\mathcal{D}_\mu\Phi_{+I}^\dagger
   \equiv(\mathcal{D}_\mu\Phi_{+I})^\dagger
   =\partial_\mu\Phi_{+I}^\dagger
   -i\Phi_{+I}^\dagger\mathcal{A}_\mu^\dagger.
\end{align}
The corresponding 2D field strength is defined by
\begin{equation}
   \mathcal{F}_{01}\equiv-i[\mathcal{D}_0,\mathcal{D}_1]
   =\partial_0\mathcal{A}_1-\partial_1\mathcal{A}_0
   +i[\mathcal{A}_0,\mathcal{A}_1].
\label{twoxfive}
\end{equation}

The $Q'$-transformation generated by the combination~(\ref{twoxone}) can be
obtained by setting $\xi_L=\xi_R=0$ and~$\bar\xi_R=-\bar\xi_L=-\bar\xi/\sqrt{2}$
in~Eqs.~(A.12) and~(A.19) of~Ref.~\cite{Sugino:2008yp} and removing
$\bar\xi$ from~$\delta_\xi$. For the gauge multiplet, we have
\begin{align}
   &Q'\mathcal{A}_\mu=0,
\nonumber\\
   &Q'\mathcal{A}_\mu^\dagger=2\psi_\mu',\qquad Q'\psi_\mu'=0,
\nonumber\\
   &Q'\chi'=-i\mathcal{F}_{01},
\nonumber\\
   &Q'\eta'=D',\qquad Q'D'=0.
\label{twoxsix}
\end{align}
Note that the complexified gauge potential, $\mathcal{A}_\mu=A_\mu-iX_\mu$, is
$Q'$-invariant. The nilpotency of~$Q'$, $(Q')^2=0$, on the gauge multiplet is
then obvious. For the fundamental matter multiplets (by using the same notation
as Ref.~\cite{Sugino:2008yp}; $\tilde m_{+I}$ are the twisted
masses~\cite{Hanany:1997vm}), we have
\begin{align}
   &Q'\phi_{+I}=0,
\nonumber\\
   &Q'\psi_{+IR}=2\mathcal{D}_{\bar z}\phi_{+I}
   +\tilde m_{+I}^*\phi_{+I},
\nonumber\\
   &Q'\psi_{+IL}=2\mathcal{D}_z\phi_{+I}+\tilde m_{+I}\phi_{+I},
\nonumber\\
   &Q'F_{+I}=-2\mathcal{D}_z\psi_{+IR}
   +2\mathcal{D}_{\bar z}\psi_{+IL}
   -2i\chi'\phi_{+I}
   -\tilde m_{+I}\psi_{+IR}+\tilde m_{+I}^*\psi_{+IL},
\nonumber\\
   &Q'\phi_{+I}^\dagger=-\bar\psi_{+IR}-\bar\psi_{+IL},
\nonumber\\
   &Q'\bar\psi_{+IR}=-F_{+I}^\dagger,
\nonumber\\
   &Q'\bar\psi_{+IL}=F_{+I}^\dagger,
\nonumber\\
   &Q'F_{+I}^\dagger=0,
\label{twoxseven}
\end{align}
where $\mathcal{D}_{z,\bar z}\equiv\frac{1}{2}(\mathcal{D}_0\mp i\mathcal{D}_1)$.
Recalling $Q'\mathcal{A}_\mu=0$ and~Eq.~(\ref{twoxfive}), the
nilpotency of~$Q'$ is again almost obvious. Note that the $Q'$-transformations
of $\psi_{+IR}$ and~$\psi_{+IL}$ have a symmetrical form as noted
in~Introduction. This is not the case with the A~model twist; see Eq.~(2.6)
of~Ref.~\cite{Sugino:2008yp}.

Each term in the continuum action of 2D $\mathcal{N}=(2,2)$ SQCD except the
superpotential (Eqs.~(2.2), (2.23) and the Wick rotation of~(A.17)
of~Ref.~\cite{Sugino:2008yp}) can be expressed in a $Q'$-exact form. The
2D SYM part is
\begin{align}
   S_{\text{2DSYM}}^{(E)}
   &=\frac{1}{g^2}\int d^2x\,\tr\Bigl(
   \frac{1}{2}F_{\mu\nu}F_{\mu\nu}+D_\mu\phi D_\mu\bar\phi
   +\frac{1}{4}[\phi,\bar\phi]^2-D^2
\nonumber\\
   &\qquad\qquad\qquad\quad{}
   +4\bar\lambda_R D_z\lambda_R
   +4\bar\lambda_L D_{\bar z}\lambda_L
   +2\bar\lambda_R[\bar\phi,\lambda_L]
   +2\bar\lambda_L[\phi,\lambda_R]\Bigr)
\nonumber\\
   &=Q'\frac{1}{g^2}\int d^2x\,\tr\left[
   -\eta'\left(D'+2D_\mu X_\mu\right)
   +i\chi'\mathcal{F}_{01}^\dagger\right].
\label{twoxeight}
\end{align}
The FI term and the theta term are
\begin{align}
   S_{\text{FI},\vartheta}^{(E)}
   &=\int d^2x\,\tr\left(\kappa D-i\frac{\vartheta}{2\pi}F_{01}\right)
\nonumber\\
   &=Q'\int d^2x\,
   \tr\left(\kappa\eta'+\frac{\vartheta}{2\pi}\chi'\right).
\label{twoxnine}
\end{align}
The action of the fundamental matter multiplets with the twisted mass
terms,\footnote{
To confirm this expression, it is useful to note
\begin{equation}
   \left(\mathcal{D}_\mu\Phi_{+I}^\dagger\right)\Phi_{+I}
   +\Phi_{+I}^\dagger\left(\mathcal{D}_\mu\Phi_{+I}\right)
   =\partial_\mu
   \left(\Phi_{+I}^\dagger\Phi_{+I}\right)+2\Phi_{+I}^\dagger X_\mu\Phi_{+I},
\end{equation}
and
\begin{equation}
   Q'\mathcal{D}_\mu\phi_{+I}^\dagger
   =-\mathcal{D}_\mu\left(\bar\psi_{+IR}+\bar\psi_{+IL}\right)
   -2i\phi_{+I}^\dagger\psi_\mu'.
\end{equation}
}
\begin{align}
   S_{\text{mat},+\tilde m}^{(E)}
   &=\int d^2x\,\sum_{I=1}^N
   \biggl[D_\mu\phi_{+I}^\dagger D_\mu\phi_{+I}
   +\frac{1}{2}\phi_{+I}^\dagger
   \{\phi-\tilde m_{+I},\bar\phi-\tilde m_{+I}^*\}\phi_{+I}
\nonumber\\
   &\qquad\qquad\qquad{}
   -F_{+I}^\dagger F_{+I}-\phi_{+I}^\dagger D\phi_{+I}
\nonumber\\
   &\qquad\qquad\qquad{}
   +2\bar\psi_{+IR}D_z\psi_{+IR}+2\bar\psi_{+IL}D_{\bar z}\psi_{+IL}
\nonumber\\
   &\qquad\qquad\qquad{}
   +\bar\psi_{+IR}(\bar\phi-\tilde m_{+I}^*)\psi_{+IL}
   +\bar\psi_{+IL}(\phi-\tilde m_{+I})\psi_{+IR}
\nonumber\\
   &\qquad{}
   -i\sqrt{2}\left(
   \phi_{+I}^\dagger(\lambda_L\psi_{+IR}-\lambda_R\psi_{+IL})
   +(-\bar\psi_{+IR}\bar\lambda_L+\bar\psi_{+IL}\bar\lambda_R)\phi_{+I}
   \right)\biggr]
\nonumber\\
   &=Q'\int d^2x\,\sum_{I=1}^N
   \biggl[
   \left(\mathcal{D}_z\phi_{+I}^\dagger\right)\psi_{+IR}
   +\left(\mathcal{D}_{\bar z}\phi_{+I}^\dagger\right)\psi_{+IL}
\nonumber\\
   &\qquad\qquad\qquad\quad{}
   +\frac{1}{2}\left(\bar\psi_{+IR}-\bar\psi_{+IL}\right)F_{+I}
   -\phi_{+I}^\dagger\eta'\phi_{+I}
\nonumber\\
   &\qquad\qquad\qquad\quad{}+\frac{1}{2}\phi_{+I}^\dagger
   \left(\tilde m_{+I}\psi_{+IR}+\tilde m_{+I}^*\psi_{+IL}\right)
   \biggr].
\end{align}
On the basis of these representations of the continuum theory with the B~model
twist, we construct a lattice formulation in the next section.

\section{Lattice model}
\subsection{Dynamical variables and lattice $Q'$-transformation}
We start with to define a lattice analogue of the $Q'$-transformation. For the
gauge multiplet, we define, on the analogy of Eq.~(\ref{twoxsix}),
\begin{align}
   &Q'U_\mu(x)=i\psi_\mu'(x)U_\mu(x),\qquad
   Q'\psi_\mu'(x)=i\psi_\mu'(x)\psi_\mu'(x),
\nonumber\\
   &Q'V_\mu(x)=i\psi_\mu'(x)V_\mu(x),
\nonumber\\
   &Q'\chi'(x)=1-U_{01}(V^{-1}U)(x),
\nonumber\\
   &Q'\eta'(x)=D'(x),\qquad Q'D'(x)=0.
\label{threexone}
\end{align}
Here, $U_\mu(x)\in U(k)$ are standard compact gauge link variables. The above
definition of $Q'$ on $U_\mu(x)$, that is exactly nilpotent, is suggested from
a naive correspondence with the continuum field, $U_\mu(x)\sim e^{iaA_\mu}$,
where $a$ denotes the lattice spacing. On the other hand, $V_\mu(x)$ are
$k\times k$ hermitian positive (noncompact) matrices corresponding to the
continuum scalar fields~$X_\mu$, $V_\mu(x)\sim e^{-aX_\mu}$. More specifically,
we introduce hermitian lattice variables~$X_\mu(x)$ and
define\footnote{Throughout this paper, as Ref.~\cite{Sugino:2008yp}, all
lattice field variables are taken to be dimensionless. These are related to
dimensionful continuum fields as,
$X_\mu(x)\to aX_\mu(x)$,
$\psi_\mu'(x)\to a^{3/2}\psi_\mu'(x)$, $\chi'(x)\to a^{3/2}\chi'(x)$,
$\eta'(x)\to a^{3/2}\eta'(x)$, $D'(x)\to a^2 D'(x)$,
$\psi_{+I,L/R}(x)\to a^{1/2}\psi_{+I,L/R}(x)$,
$\bar\psi_{+I,L/R}(x)\to a^{1/2}\bar\psi_{+I,L/R}(x)$,
$F_{+I}(x)\to aF_{+I}(x)$, $F_{+I}(x)^\dagger\to aF_{+I}(x)^\dagger$,
where all fields in the right-hand sides are continuum ones.
}
\begin{equation}
   V_\mu(x)\equiv e^{-X_\mu(x)}.
\label{threextwo}
\end{equation}
$U_{\mu\nu}(x)$ denote the standard plaquette variables
\begin{equation}
   U_{\mu\nu}(x)\equiv U_{\mu\nu}(U)(x)\equiv
   U_\mu(x)U_\nu(x+a\hat\mu)U_\mu(x+a\hat\nu)^{-1}U_\nu(x)^{-1}
\end{equation}
and $U_{01}(V^{-1}U)(x)$ in Eq.~(\ref{threexone}) is defined by the substitution
$U_\mu(x)\to V_\mu(x)^{-1}U_\mu(x)$ in this expression. Note that in general
$U_{01}(V^{-1}U)(x)$ are not unitary; they are elements of~$GL(k,\mathbb{C})$.

We assume that $V_\mu(x)$ (for both $\mu=0$ and~1) are \emph{site\/} variables
transforming as adjoint under lattice gauge transformations at the point~$x$
(the same is assumed for $X_\mu(x)$, $\psi_\mu'(x)$, $\chi'(x)$, $\eta'(x)$
and~$D'(x)$). Then the both sides of each relation of~Eq.~(\ref{threexone})
have the same gauge transformation property regarding $Q'$ as a gauge singlet.

With the above naive correspondence, $U_\mu(x)\sim e^{iaA_\mu}$
and~$V_\mu(x)\sim e^{-aX_\mu}$, the combination $V_\mu(x)^{-1}U_\mu(x)$ would
correspond to the exponential of the complexified gauge potential
$\mathcal{A}_\mu=A_\mu-iX_\mu$,
$V_\mu(x)^{-1}U_\mu(x)\sim e^{ia\mathcal{A}_\mu}$. In fact, the
combination~$V_\mu(x)^{-1}U_\mu(x)$ is invariant under lattice
$Q'$-transformation~(\ref{threexone}). Noting this, it is easy to see that
lattice $Q'$-transformation~(\ref{threexone}) is nilpotent $(Q')^2=0$ on the
gauge multiplet.

For the matter multiplets in the fundamental representation, we set
\begin{align}
   &Q'\phi_{+I}(x)=0,
\nonumber\\
   &Q'\psi_{+IR}(x)=2a\mathcal{D}_{\bar z}\phi_{+I}(x)
   +a\tilde m_{+I}^*\phi_{+I}(x),
\nonumber\\
   &Q'\psi_{+IL}(x)
   =2a\mathcal{D}_z\phi_{+I}(x)+a\tilde m_{+I}\phi_{+I}(x),
\nonumber\\
   &Q'F_{+I}(x)=-2a\mathcal{D}_z\psi_{+IR}(x)
   +2a\mathcal{D}_{\bar z}\psi_{+IL}(x)
\nonumber\\
   &\qquad\qquad\qquad{}
   -2i\chi'(x)V_1(x)^{-1}U_1(x)V_0(x+a\hat1)^{-1}U_0(x+a\hat1)
   \phi_{+I}(x+a\hat0+a\hat1)
\nonumber\\
   &\qquad\qquad\qquad{}
   -a\tilde m_{+I}\psi_{+IR}(x)+a\tilde m_{+I}^*\psi_{+IL}(x),
\nonumber\\
   &Q'\phi_{+I}(x)^\dagger=-\bar\psi_{+IR}(x)-\bar\psi_{+IL}(x),
\nonumber\\
   &Q'\bar\psi_{+IR}(x)=-F_{+I}(x)^\dagger,
\nonumber\\
   &Q'\bar\psi_{+IL}(x)=F_{+I}(x)^\dagger,
\nonumber\\
   &Q'F_{+I}(x)^\dagger=0.
\label{threexfour}
\end{align}
For the covariant differences for a generic lattice field $\Phi_{+I}(x)$ in the
fundamental representation, we adopt the forward differences
\begin{align}
   &a\mathcal{D}_\mu\Phi_{+I}(x)
   \equiv V_\mu(x)^{-1}U_\mu(x)\Phi_{+I}(x+a\hat\mu)-\Phi_{+I}(x),
\nonumber\\
   &a\mathcal{D}_\mu\Phi_{+I}(x)^\dagger
   \equiv\left(a\mathcal{D}_\mu\Phi_{+I}(x)\right)^\dagger
   =\Phi_{+I}(x+a\hat\mu)^\dagger U_\mu(x)^{-1}V_\mu(x)^{-1}
   -\Phi_{+I}(x)^\dagger
\label{threexfive}
\end{align}
and, as in the continuum theory, $\mathcal{D}_{z,\bar z}
\equiv\frac{1}{2}(\mathcal{D}_0\mp i\mathcal{D}_1)$.
The nilpotency of~$Q'$ in~Eq.~(\ref{threexfour}) is almost obvious except that
for the auxiliary field $F_{+I}(x)$. This nilpotency $(Q')^2F_{+I}(x)=0$ follows
from the identity
\begin{align}
   &\left(1-U_{01}(V^{-1}U)(x)\right)
   V_1(x)^{-1}U_1(x)V_0(x+a\hat1)^{-1}U_0(x+a\hat1)\phi_{+I}(x+a\hat0+a\hat1)
\nonumber\\
   &=2ia^2[\mathcal{D}_z,\mathcal{D}_{\bar z}]\phi_{+I}(x),
\end{align}
that is a lattice analogue of the relation~(\ref{twoxfive}).

We thus defined $Q'$-transformation on the lattice that is completely nilpotent
on all lattice variables. Note that the nilpotency $(Q')^2=0$ holds without
referring to any equivalence under the gauge or flavor rotations; this is quite
different from the cases in the A~model
twist~\cite{Sugino:2008yp,Kikukawa:2008xw}.

\subsection{Lattice action}
Next, we define the lattice action. The SYM part is defined by, on the analogy
of Eq.~(\ref{twoxeight}),
\begin{equation}
   S_{\text{2DSYM}}^{\text{LAT}}
   =Q'\frac{1}{a^2g^2}\sum_x\tr\left[
   -\eta'(x)\left(D'(x)+W(x)\right)
   +\chi'(x)\left(1-U_{01}(V^{-1}U)(x)\right)^\dagger\right].
\end{equation}
As a possible choice of $W(x)$, that is a lattice counterpart
of~$2D_\mu X_\mu$, we take
\begin{equation}
   W(x)\equiv2\sum_\mu\left[
   V_\mu(x)^{-1}+U_\mu(x-a\hat\mu)^{-1}V_\mu(x-a\hat\mu)U_\mu(x-a\hat\mu)-2
   \right]
\label{threexeight}
\end{equation}
and this in fact reduces to $2a^2D_\mu X_\mu$ in the naive continuum limit,
$U_\mu(x)\sim e^{iaA_\mu}$, $V_\mu(x)\sim e^{-aX_\mu}$ and~$a\to0$. Note that
$W(x)$ are hermitian matrices.

The FI term and the theta term are defined by
(see~Eq.~(\ref{twoxnine}))\footnote{This theta term is not topologically
invariant with finite lattice spacings. One could instead use
$S_\vartheta^{\text{LAT}}=-\vartheta/(2\pi)\sum_x\tr\ln U_{01}(U)(x)$,
that is topologically invariant if configurations with an eigenvalue of
$U_{01}(U)(x)$ being $-1$ for a certain~$x$ are excised
(see~Ref.~\cite{Fujiwara:2000wn} and references cited therein), as precisely
the case when the admissibility is
imposed~\cite{Sugino:2008yp,Kikukawa:2008xw}. However, even without imposing
the admissibility, such configurations should not contribute to functional
integrals in the continuum limit (see Sec.~\ref{sec:threexthree}) and this term
would practically work as a topological (and thus $Q'$) invariant in the
continuum limit.}
\begin{equation}
   S_{\text{FI},\vartheta}^{\text{LAT}}
   =Q'\sum_x\tr\left[
   \kappa\eta'(x)+\frac{\vartheta}{2\pi}\chi'(x)\right].
\end{equation}

The lattice action for matter multiplets is almost the same as the continuum
one:
\begin{align}
   S_{\text{mat},+\tilde m}^{\text{LAT}}&=Q'\sum_x\sum_{I=1}^N
   \biggl[
   \left(a\mathcal{D}_z\phi_{+I}(x)^\dagger\right)\psi_{+IR}(x)
   +\left(a\mathcal{D}_{\bar z}\phi_{+I}(x)^\dagger\right)
   \psi_{+IL}(x)
\nonumber\\
   &\qquad\qquad\qquad{}
   +\frac{1}{2}\left(\bar\psi_{+IR}(x)-\bar\psi_{+IL}(x)\right)F_{+I}(x)
   -\phi_{+I}(x)^\dagger\eta'(x)\phi_{+I}(x)
\nonumber\\
   &\qquad\qquad\qquad{}+\frac{1}{2}\phi_{+I}(x)^\dagger
   \left(a\tilde m_{+I}\psi_{+IR}(x)+a\tilde m_{+I}^*\psi_{+IL}(x)\right)
   \biggr].
\label{threexten}
\end{align}
Note, however, that the lattice covariant differences appearing in these
expressions are forward ones~(\ref{threexfive}).

From the above construction and from the nilpotency~$(Q')^2=0$, it is clear
that our lattice action is invariant under gauge and fermionic
$Q'$~transformations.

The total lattice action possesses also some global symmetries; one is the
$U(1)_V$ symmetry, under which
\begin{align}
   &\psi_\mu'(x)\to e^{i\alpha}\psi_\mu'(x),\qquad
   \chi'(x)\to e^{-i\alpha}\chi'(x),\qquad
   \eta'(x)\to e^{-i\alpha}\eta'(x),
\nonumber\\
   &\psi_{+I,L/R}(x)\to e^{-i\alpha}\psi_{+I,L/R}(x),\qquad
   \bar\psi_{+I,L/R}(x)\to e^{i\alpha}\bar\psi_{+I,L/R}(x),
\nonumber\\
   &F_{+I}(x)\to e^{-2i\alpha}F_{+I}(x),\qquad
   F_{+I}(x)^\dagger\to e^{2i\alpha}F_{+I}(x)^\dagger,
\label{threexeleven}
\end{align}
and other variables are kept intact. Another is $U(1)^N$ symmetry that rotates
each fundamental multiplet independently\footnote{If some twisted masses are
degenerated, this symmetry enhances accordingly.}
\begin{equation}
   \Phi_{+I}(x)\to e^{i\alpha_{+I}}\Phi_{+I}(x),\qquad
   \Phi_{+I}(x)^\dagger\to e^{-i\alpha_{+I}}\Phi_{+I}(x)^\dagger.
\label{threextwelve}
\end{equation}
Besides symmetry under discrete translations by the lattice unit, the present
lattice action does not possess further (fermionic as well as bosonic)
symmetries that were present in the continuum theory.

\subsection{Weak coupling saddle point of the lattice action}
\label{sec:threexthree}
In the naive continuum limit~$a\to0$, in which one assumes
$U_\mu(x)\sim e^{iaA_\mu}$ and~$V_\mu(x)\sim e^{-aX_\mu}$, our lattice action
reproduces the continuum action of 2D $\mathcal{N}=(2,2)$ $U(k)$ SQCD with
$N$~fundamental matter multiplets. A perturbative argument
in~Sec.~\ref{sec:threexsix} then indicates that the continuum limit of the
present lattice model is given by the weak coupling limit~$\beta\to\infty$,
where $1/(a^2g^2)\equiv\beta/(2k)$,
$\kappa=N/(4\pi)\ln\beta+\text{const.}$,
$a\tilde m_{+I}=(\tilde m_{+I}/g)\sqrt{2k/\beta}$
and~$a\tilde m_{+I}^*=(\tilde m_{+I}^*/g)\sqrt{2k/\beta}$.

However, for the above perturbative picture on the basis of expansion around
$U_\mu(x)=1$ and~$V_\mu(x)=1$ to be consistent, the configuration $U_\mu(x)=1$
and~$V_\mu(x)=1$ (up to gauge transformations) should give the unique saddle
point in the weak coupling limit. It suffices if this is so for an infinite
lattice. Whether this is really the case or not, however, could generally be a
nontrivial issue. In fact,
in~Refs.~\cite{Sugino:2004qd,Sugino:2006uf,Sugino:2008yp,Kikukawa:2008xw},
the admissibility~\cite{Luscher:1998du} was incorporated in the lattice action
to remove weak coupling saddle points that have no continuum counterpart.

In the present lattice model, at least for $\vartheta=0$, we can see that the
unique saddle point in the weak coupling limit for an infinite lattice is
$U_\mu(x)=1$ and~$V_\mu(x)=1$ up to gauge transformations. The argument proceeds
as follows.\footnote{In usual lattice gauge theory with compact gauge link
variables, such as lattice QCD, the weak coupling saddle point is not affected
by the presence of fermions, because the fermion determinant would be a bounded
function of link variables and consequently it cannot modify saddle points
for~$\beta\to\infty$. Strictly speaking, this reasoning cannot be applied to
our present system because the fermion determinant could be an unbounded
function of \emph{noncompact\/} scalar fields. The fermion determinant in
principle could balance with bosonic action~(\ref{threexthirteen}) and modify
the saddle points for~$\beta\to\infty$. We do not consider this possibility
below because this could occur only at the ``boundary'' of the field space,
such as $V_\mu=0$ or~$V_\mu=+\infty$, and if this occurs, our lattice
formulation would be meaningless in any case.}

We first seek saddle points for~$\beta\to\infty$ on a lattice with a finite
number of lattice points, $N_\mu$ in the $\mu$-direction, to avoid a subtlety
associated with an infinite lattice. We assume periodic boundary conditions for
bosonic fields. After obtaining all saddle points on this finite lattice, we
send $N_\mu$ to infinity yielding saddle points for an infinite lattice.

The bosonic part of the lattice action, after integrating over the auxiliary
fields, takes the form
\begin{align}
   &S_{\text{2DSYM}}^{\text{LAT}}+S_{\text{FI},\vartheta}^{\text{LAT}}
   +S_{\text{mat},+\tilde m}^{\text{LAT}}
\nonumber\\
   &=\frac{\beta}{2k}\sum_x\tr\Biggl[
   \frac{1}{4}\left\{W(x)+\frac{2k}{\beta}\left(
   \sum_{I=1}^N\phi_{+I}(x)\phi_{+I}(x)^\dagger-\kappa\right)\right\}^2
\nonumber\\
   &\qquad\qquad\qquad{}
   +\left(1-U_{01}(V^{-1}U)(x)\right)
   \left(1-U_{01}(V^{-1}U)(x)+\frac{2k}{\beta}\frac{\vartheta}{2\pi}
   \right)^\dagger\Biggr]
\nonumber\\
   &\quad{}
   +\sum_x\sum_{I=1}^N
   \left(a\mathcal{D}_\mu\phi_{+I}(x)
   +a\tilde m_{+I,\mu}\phi_{+I}(x)
   \right)^\dagger
   \left(a\mathcal{D}_\mu\phi_{+I}(x)
   +a\tilde m_{+I,\mu}\phi_{+I}(x)
   \right),
\label{threexthirteen}
\end{align}
where
\begin{equation}
   \tilde m_{+I,\mu}\equiv\begin{cases}
   \Real\tilde m_{+I},&\text{for $\mu=0$},\\
   -\Imag\tilde m_{+I},&\text{for $\mu=1$},
   \end{cases}
\end{equation}
and we have assumed that $\tilde m_{+I}$ and $\tilde m_{+I}^*$ are complex
conjugate to each other. Therefore, after integrating over the auxiliary
fields, the bosonic part of the lattice action is real and positive
semi-definite for $\vartheta=0$ (recall that $W(x)$ are hermitian); this is
certainly a desired property. From the above expression, for $\vartheta=0$,
saddle points for~$\beta\to\infty$ are specified by
\begin{align}
   &W(x)+\frac{2k}{\beta}\sum_{I=1}^N\phi_{+I}(x)\phi_{+I}(x)^\dagger
   =\frac{2k}{\beta}\kappa,
\label{threexfifteen}
\\
   &U_{01}(V^{-1}U)(x)=1.
\label{threexsixteen}
\end{align}
We then apply $\sum_x\tr$ to both sides of Eq.~(\ref{threexfifteen}) to yield
\begin{equation}
   \sum_x\tr W(x)
   +\frac{2k}{\beta}\sum_x\sum_{I=1}^N\phi_{+I}(x)^\dagger\phi_{+I}(x)
   =\frac{2k}{\beta}\kappa\sum_x\tr1.
\label{threexseventeen}
\end{equation}
We further note that $\sum_x\tr W(x)$ is positive semi-definite
\begin{align}
   \sum_x\tr W(x)
   &=\sum_x2\sum_\mu\tr\left[V_\mu(x)^{-1}+V_\mu(x-a\hat\mu)-2\right]
\nonumber\\
   &=\sum_x2\sum_\mu\tr\left[V_\mu(x)^{-1}+V_\mu(x)-2\right]
\nonumber\\
   &=\sum_x2\sum_\mu\tr\left[\left(V_\mu(x)^{-1/2}-V_\mu(x)^{1/2}\right)^2\right]
   \geq0
\label{threexeighteen}
\end{align}
(recall that $V_\mu(x)$ are positive and thus the square root can always be
defined).

Now, in the $\beta\to\infty$ limit, the right-hand side
of~Eq.~(\ref{threexseventeen}) vanishes because $\kappa$ grows at most
$\sim\ln\beta$. In the left-hand side of~Eq.~(\ref{threexseventeen}), there
cannot occur cancellation between the first and the second terms because both
are positive semi-definite. These imply that $\sum_x\tr W(x)=0$
for~$\beta\to\infty$. Then, from~Eq.~(\ref{threexeighteen}), we have
$V_\mu(x)=1$ at saddle points for~$\beta\to\infty$. Plugging this $V_\mu(x)=1$
into~Eq.~(\ref{threexsixteen}), we see that at saddle points, the gauge
plaquette is unity~$U_{01}(U)(x)=1$. This is an identical condition for the
weak coupling saddle point with the \emph{standard plaquette action}. As shown
in~Appendix~\ref{app:a}, the most general form of such flat connections
satisfying~$U_{01}(U)(x)=1$ on a periodic lattice is given
by~$U_\mu(x)=g(x)T_\mu g(x+a\hat\mu)^{-1}$, where the gauge
transformation~$g(x)\in U(k)$ is periodic on the lattice and the constant
factor~$T_\mu$ is given by~Eq.~(\ref{axfour}). Therefore, up to gauge
transformations, the weak coupling saddle points are given
by~$U_\mu(x)=T_\mu$ and~$V_\mu(x)=1$.

Finally, in the infinite lattice limit $N_\mu\to\infty$, $T_\mu\to1$ as
Eq.~(\ref{axfour}) shows. Therefore, the saddle point for~$\beta\to\infty$ on
an infinite lattice is given by $U_\mu(x)=1$ and~$V_\mu(x)=1$ up to gauge
transformations. This completes our argument for the weak coupling saddle
point.\footnote{If one wishes, it \emph{is\/} possible to tailor a
$W(x)$ such that there are infinite (gauge inequivalent) weak coupling saddle
points for the scalar fields~$V_\mu(x)$. For example, with the choice
\begin{align}
   W(x)\equiv-2\sum_\mu\left[V_\mu(x)
   -U_\mu(x-a\hat\mu)^{-1}V_\mu(x-a\hat\mu)U_\mu(x-a\hat\mu)\right]
   +\frac{1}{k}L(x)
\end{align}
and
\begin{equation}
   L(x)=\tr\left[2-U_{01}(U)(x)-U_{01}(U)(x)^{-1}
   +\sum_\mu a^2D_\mu\phi(x)D_\mu\bar\phi(x)
   +\frac{1}{4}[\phi(x),\bar\phi(x)]^2\right]
\end{equation}
($\phi$ and~$\bar\phi$ are given by Eq.~(\ref{twoxtwo})
and~$X_\mu(x)\equiv-\ln V_\mu(x)$; $D_\mu$ are forward covariant differences
for adjoint fields with gauge link variables $U_\mu(x)$ used), by an argument
similar to that in the main text, one sees that the saddle points (for an
infinite lattice) are given by $U_\mu(x)=1$ and~$X_\mu(x)=\text{const.}$
and~$[X_0,X_1]=0$ up to gauge transformations. These configurations (with
$\phi_{+I}(x)=0$) provide also a noncompact set of zeros of the lattice bosonic
action when~$\kappa=0$, corresponding to the Coulomb branch in the continuum
theory. The free kinetic terms of fermions resulting from the above~$W(x)$ are
identical to those in the main text and thus this choice does not lead to
the species doubling.}

In the present lattice model on the finite lattice, moreover, one can see that
the space of zeros of the bosonic action is, if it is not empty, always compact
even for finite~$\beta$ and~$\kappa$; this follows
from~Eq.~(\ref{threexseventeen}), the condition that such zeros must satisfy.
In Eq.~(\ref{threexseventeen}), the first term defines the square of a distance
between $V_\mu(x)$ and $V_\mu(x)\equiv1$ and similarly the second term defines a
norm of $\phi_{+I}$, both are positive semi-definite. It is then obvious that
any solution of the relation~(\ref{threexseventeen}) cannot grows indefinitely
because the right-hand side remains finite for any nonzero~$\beta$; this shows
that the space of zeros of the bosonic action is compact.\footnote{This
property is shared also by lattice formulations of 2D $\mathcal{N}=(2,2)$ SYM
in~Refs.~\cite{Sugino:2006uf} and~\cite{Suzuki:2005dx} which use
\emph{compact\/} lattice scalar fields. The former formulation possesses a
manifest fermionic symmetry.}

\subsection{Absence of the species doubling}
Once the expansion around $U_\mu(x)=1$ and~$V_\mu(x)=1$ is justified, it is
straightforward to see that the present lattice formulation is free from the
species doubling. Setting $U_\mu(x)=V_\mu(x)=1$, we have
$S_{\text{2DSYM}}^{\text{LAT}}=-2/(a^2g^2)\*\sum_x\tr[\bar\psi(x)aD\psi(x)]$,
where $\bar\psi\equiv(\bar\lambda_L,\bar\lambda_R)$
and~$\psi^T\equiv(\lambda_R,\lambda_L)$, and
\begin{equation}
   aD\equiv
   \sum_\mu\gamma_\mu\frac{1}{2}\left(a\partial_\mu+a\partial_\mu^*\right)
   -\frac{1}{2}
   \left(a^2\partial_0^*\partial_0-i\gamma_5a^2\partial_1^*\partial_1\right).
\label{threextwentyone}
\end{equation}
In this expression,
$\gamma_0\equiv\bigl(\begin{smallmatrix}0&-1\\-1&0\end{smallmatrix}\bigr)$,
$\gamma_1\equiv\bigl(\begin{smallmatrix}0&-i\\i&0\end{smallmatrix}\bigr)$ and
$\gamma_5\equiv\bigl(\begin{smallmatrix}1&0\\0&-1\end{smallmatrix}\bigr)$,
and $\partial_\mu$ and~$\partial_\mu^*$ are the forward and the backward
difference operators, respectively. The second term
in~Eq.~(\ref{threextwentyone}) acts as a Wilson term and, since
$a^2D^\dagger D=-\sum_\mu a^2\partial_\mu^*\partial_\mu$, the free Dirac
operator~$D$ vanishes only at the origin of the Brillouin zone. That is, there
is no species doubling.

Similarly, for the matter sector, for $U_\mu(x)=V_\mu(x)=1$
and~$\phi_{+I}(x)=0$, we have
\begin{equation}
   S_{\text{mat},+\tilde m}^{\text{LAT}}
   =\sum_x\sum_{I=1}^N
   \bar\psi_{+I}\gamma_5\left(aD
   -a\tilde m_{+I}\frac{1+\gamma_5}{2}
   -a\tilde m_{+I}^*\frac{1-\gamma_5}{2}
   \right)\gamma_5\psi_{+I},
\end{equation}
where $\bar\psi_{+I}\equiv(\bar\psi_{+IL},\bar\psi_{+IR})$
and~$\psi_{+I}^T\equiv(\psi_{+IR},\psi_{+IL})$. This kinetic
operator reproduces the correct dispersion relation for massive fermions near
the origin of the Brillouin zone.

\subsection{Invariant integration measure}
For our lattice formulation to be invariant under gauge, $Q'$, $U(1)_V$
and~$U(1)^N$ transformations, not only the lattice action but also the
integration measure must be invariant under these transformations.
Except for the scalar fields~$V_\mu(x)$, the integration measure is standard:
\begin{equation}
   \prod_x\left[\prod_{\mu=0}^1dU_\mu(x)\right]
   \prod_{\mathtt{A}}d\psi_0^{\prime\mathtt{A}}(x)\,d\psi_1^{\prime\mathtt{A}}(x)
   \,d\chi^{\prime\mathtt{A}}(x)\,d\eta^{\prime\mathtt{A}}(x)
   \,dD^{\prime\mathtt{A}}(x)
   \left[\prod_{I=1}^N(d\mu_{\text{mat},+I})\right],
\label{threextwentythree}
\end{equation}
where
\begin{align}
   &(d\mu_{\text{mat},+I})
\nonumber\\
   &\equiv
   \prod_x\prod_{i=1}^kd\phi_{+Ii}(x)\,d\phi_{+Ii}(x)^*
   \,d\psi_{+ILi}(x)\,d\psi_{+IRi}(x)
   \,d\bar\psi_{+ILi}(x)\,d\bar\psi_{+IRi}(x)
\nonumber\\
   &\qquad\qquad\qquad\qquad\qquad{}\times
   dF_{+Ii}(x)\,dF_{+Ii}(x)^*.
\end{align}
In Eq.~(\ref{threextwentythree}), $dU_\mu(x)$ is the conventional Haar measure
on~$U(k)$. It can be seen that the above measure is invariant under gauge,
$Q'$, $U(1)_V$ and~$U(1)^N$ transformations. The argument is essentially the
same as that of Ref.~\cite{Sugino:2006uf}. In particular, $Q'$-invariance of
the Haar measure follows from the fact that the $Q'$~transformation on link
variables $U_\mu(x)$ can be regarded as a left-multiplication of a group
element $U_\mu(x)\to g(x)U_\mu(x)$, where $g(x)\in U(k)$, as shown in~Eq.~(2.3)
of~Ref.~\cite{Sugino:2006uf}.

On the other hand, the definition of an invariant integration measure for the
scalar fields $V_\mu(x)$ is somewhat intricate. We start with the following
norm of a variation of~$V_\mu(x)$
\begin{equation}
   \left\|\delta V_\mu(x)\right\|^2
   \equiv\tr\left[V_\mu(x)^{-1}\delta V_\mu(x)V_\mu(x)^{-1}\delta V_\mu(x)\right].
\label{threextwentyfive}
\end{equation}
This norm is positive semi-definite, because $\|\delta V_\mu(x)\|^2
=\tr[(V_\mu(x)^{-1/2}\delta V_\mu(x)V_\mu(x)^{-1/2})^2]$. An integration measure
associated with this norm, according to a standard recipe, is given by
\begin{equation}
   \prod_x\prod_\mu\left[\prod_{\mathtt{A}}dV_\mu^{\mathtt{A}}(x)\right]
   \sqrt{\det_{\mathtt{A},\mathtt{B}}
   \tr\left[V_\mu(x)^{-1}T^{\mathtt{A}}V_\mu(x)^{-1}T^{\mathtt{B}}\right]}.
\label{threextwentysix}
\end{equation}
The point is that the norm~(\ref{threextwentyfive}) is invariant under the
substitutions (I)~$V_\mu(x)\to u_\mu(x)V_\mu(x)u_\mu(x)^{-1}$
and~(II)~$V_\mu(x)\to h_\mu(x)V_\mu(x)h_\mu(x)$, where $u_\mu(x)$ are unitary
matrices and $h_\mu(x)$ are invertible hermitian matrices. From this
invariance of the norm, it follows that also the
measure~(\ref{threextwentysix}) is invariant under these substitutions, as can
be verified explicitly
by using $\sum_{\mathtt{A}}(T^{\mathtt{A}})_{ij}(T^{\mathtt{A}})_{kl}
=(1/2)\delta_{il}\delta_{jk}$.
From the invariance under~(I), gauge invariance of the measure is obvious
because gauge transformations take the form of~(I). Furthermore, the measure is
invariant also under the $Q'$-transformation as follows.

The measure~(\ref{threextwentysix}) is of course invariant under an
infinitesimal version of the substitutions,
(i)~$V_\mu(x)\to V_\mu(x)+i\theta_\mu^{\mathtt{A}}(x)[T^{\mathtt{A}},V_\mu(x)]$
and~(ii)~$V_\mu(x)\to V_\mu(x)
+\zeta_\mu^{\mathtt{A}}(x)\{T^{\mathtt{A}},V_\mu(x)\}$,
where $\theta_\mu^{\mathtt{A}}(x)$ and $\zeta_\mu^{\mathtt{A}}(x)$ are
infinitesimal real parameters. However, this invariance holds even if we regard
$\theta_\mu^{\mathtt{A}}(x)$ and $\zeta_\mu^{\mathtt{A}}(x)$ as infinitesimal
\emph{complex\/} parameters, because no complex conjugation is involved for the
invariance. In particular, we may set
$\theta_\mu^{\mathtt{A}}(x)=(1/2)\bar\xi\psi_\mu^{\prime\mathtt{A}}(x)$
and~$\zeta_\mu^{\mathtt{A}}(x)=(i/2)\bar\xi\psi_\mu^{\prime\mathtt{A}}(x)$, where
$\bar\xi$ is a Grassmann parameter. Then a combination of the above (i)
and~(ii) becomes $V_\mu(x)\to V_\mu(x)+i\bar\xi\psi_\mu'(x)V_\mu(x)$, the
$Q'$-transformation on $V_\mu(x)$. This shows that the
measure~(\ref{threextwentysix}) is invariant also under the
$Q'$-transformation. Thus, the above defined measure~(\ref{threextwentysix})
has desired invariance properties.

However, if integration variables are~$V_\mu(x)$, one has to take into account
the fact that $V_\mu(x)$ are positive matrices. Practically, an integration
measure for hermitian variables $X_\mu(x)$ in~Eq.~(\ref{threextwo}) should be
more useful. By rewriting invariant norm~(\ref{threextwentyfive}) in terms of a
variation of~$X_\mu(x)$, we have
\begin{equation}
   \prod_x\prod_\mu\left[\prod_{\mathtt{A}}dX_\mu^{\mathtt{A}}(x)\right]
   \sqrt{\det_{\mathtt{A},\mathtt{B}}M_\mu^{\mathtt{AB}}(x)},
\label{threextwentyseven}
\end{equation}
where
\begin{equation}
   M_\mu^{\mathtt{AB}}(x)\equiv
   \int_0^1d\alpha\,\int_0^1d\beta\,\tr
   \left[e^{(\alpha-\beta)X_\mu(x)}T^{\mathtt{A}}
   e^{-(\alpha-\beta)X_\mu(x)}T^{\mathtt{B}}\right],
\label{threextwentyeight}
\end{equation}
and the integration region of each variable $X_\mu^{\mathtt{A}}(x)$ is
$(-\infty,+\infty)$.

From integration variables $X_\mu^{\mathtt{A}}(x)$, one can construct the
matrices~$V_\mu(x)$ by $V_\mu(x)=u_\mu(x)e^{-\lambda_\mu(x)}u_\mu(x)^{-1}$,
where $\lambda_\mu(x)\equiv\diag(\lambda_{\mu 1}(x),\dots,\lambda_{\mu k}(x))$
and $\lambda_{\mu i}(x)$ ($i=1$, 2, \dots, $k$) are eigenvalues of
$X_\mu(x)=\sum_{\mathtt{A}}X_\mu^{\mathtt{A}}(x)T^{\mathtt{A}}$; $u_\mu(x)$ are
unitary matrices that diagonalize $X_\mu(x)$,
$X_\mu(x)=u_\mu(x)\lambda_\mu(x)u_\mu(x)^{-1}$. In terms of these eigenvalues,
the volume element is expressed as
\begin{equation}
   \sqrt{\det_{\mathtt{A},\mathtt{B}}M_\mu^{\mathtt{AB}}(x)}
   =\frac{1}{\sqrt{2^k}}\prod_{i<j}
   \frac{\cosh(\lambda_{\mu i}(x)-\lambda_{\mu j}(x))-1}
   {(\lambda_{\mu i}(x)-\lambda_{\mu j}(x))^2}\geq\frac{1}{\sqrt{2^{k^2}}}.
\label{threextwentynine}
\end{equation}
One can directly confirm that the measure~(\ref{threextwentyseven})
with~Eq.~(\ref{threextwentynine}) is invariant under the above substitutions
(i) and~(ii), and thus under the $Q'$~transformation.

Also, in the hybrid Monte Carlo algorithm, one needs to compute variations
of~$V_\mu(x)$ and of the volume element with respect to the integration
variables~$X_\mu^{\mathtt{A}}(x)$. They are given by
\begin{align}
   &\left(\delta V_\mu(x)\right)_{ij}
\nonumber\\
   &=\sum_{k,l}\left(u_\mu(x)\right)_{ik}
   \frac{e^{-\lambda_{\mu k}(x)}-e^{-\lambda_{\mu l}(x)}}
   {\lambda_{\mu k}(x)-\lambda_{\mu l}(x)}
   \left(u_\mu(x)^{-1}T^{\mathtt{A}}u_\mu(x)\right)_{kl}
   \left(u_\mu(x)^{-1}\right)_{lj}
   \delta X_\mu^{\mathtt{A}}(x)
\end{align}
and
\begin{equation}
   \delta\ln\sqrt{\det_{\mathtt{A},\mathtt{B}}M_\mu^{\mathtt{AB}}(x)}
   =\sum_{i\neq j}f(\lambda_{\mu i}(x)-\lambda_{\mu j}(x))
   \left(u_\mu(x)^{-1}T^{\mathtt{A}}u_\mu(x)\right)_{ii}
   \delta X_\mu^{\mathtt{A}}(x),
\label{threexthirtyone}
\end{equation}
where
\begin{equation}
   f(x)\equiv\frac{\sinh x}{\cosh x-1}-\frac{2}{x}.
\label{threexthirtytwo}
\end{equation}
These expressions should be useful in actual Monte Carlo
simulations.\footnote{For gauge groups $U(1)$ and~$U(2)$, $V_\mu(x)$ and
the volume element can directly be expressed by $X_\mu(x)$: For $U(1)$,
$V_\mu(x)=e^{-X_\mu^0(x)/\sqrt{2}}$
and~$\sqrt{\det_{\mathtt{A},\mathtt{B}}M_\mu^{\mathtt{AB}}(x)}=1/\sqrt{2}$.
For $U(2)$,
\begin{equation}
   V_\mu(x)=e^{-X_\mu^0(x)/2}
   \left(\cosh(|\Vec X_\mu(x)|/2)
   -\frac{\sinh(|\Vec X_\mu(x)|/2)}
   {|\Vec X_\mu(x)|/2}\,\Vec T\cdot\Vec X_\mu(x)\right),
\end{equation}
where $\Vec T\equiv(\sigma^1/2,\sigma^2/2,\sigma^3/2)$
and~$\Vec X_\mu(x)\equiv(X_\mu^1(x),X_\mu^2(x),X_\mu^3(x))$,
and $\sqrt{\det_{\mathtt{A},\mathtt{B}}M_\mu^{\mathtt{AB}}(x)}
=[\cosh(|\Vec X_\mu(x)|)-1]/2|\Vec X_\mu(x)|^2$.}

\subsection{Continuum limit}
\label{sec:threexsix}
In the present super-renormalizable gauge theory, the continuum limit is given
by $\beta\to\infty$, where $\beta\equiv 2k/(a^2g^2)$. In this section, we argue
that (within perturbation theory) all symmetries broken by lattice
regularization are restored in the continuum limit without any fine tuning. For
this argument, it is convenient to rescale continuum matter multiplets as
$\Phi_{+I}\to(1/g)\Phi_{+I}$ so that the mass dimensions of fields in matter
multiplets become the same as the gauge multiplet (that is, bosonic
fields have mass dimension~1, fermionic have~$3/2$, the auxiliary fields~2).

Generally speaking, symmetries broken by UV regularization could be recovered
by supplementing appropriately chosen \emph{local\/} counterterms. The most
general form of local terms in the effective action, from the dimensional
consideration, is
\begin{equation}
   \left(c_0\frac{a^{p-4}}{g^2}+c_1a^{p-2}+c_2a^pg^2+\cdots\right)
   \int d^2x\,\varphi^a\partial^b\psi^{2c}\mathcal{A}^d,\qquad
   p\equiv a+b+3c+2d\geq0,
\label{threexthirtyfour}
\end{equation}
up to some powers of possible logarithmic ($\ln a$) factors. In this
expression, $\varphi$ symbolically denotes bosonic fields in the continuum
theory except the auxiliary fields, $\psi$ denotes fermionic fields, and
$\mathcal{A}$ denotes the auxiliary fields; $\partial$ a derivative.
Abbreviated terms in the parentheses are of strictly positive powers in~$a$, so
they are irrelevant in the continuum limit. The coefficients $c_0$, $c_1$
and~$c_2$ are some dimensionless combinations of the parameters $\kappa$,
$\vartheta$, $g/\tilde m_{+I}$ and~$g/\tilde m_{+I}^*$.

Now, local operators that are proportional to the first term in the parentheses
of~Eq.~(\ref{threexthirtyfour}) arise only at the tree-level approximation;
that is, from the naive continuum limit of the lattice action. Our lattice
action reproduces, in this limit, the classical action of 2D
$\mathcal{N}=(2,2)$ SQCD. Those local terms are simply the terms in the
classical action
$S_{\text{2DSYM}}^{(E)}+S_{\text{FI},\vartheta}^{(E)}+S_{\text{mat},+\tilde m}^{(E)}$.

Terms being proportional to the second term in the parentheses
of~Eq.~(\ref{threexthirtyfour}) arise at the one-loop level or lower. For them
to be relevant or marginal, we have to have~$p\leq2$. Most of possible local
operators from the dimensional grounds are excluded by the gauge invariance.
Further requiring $Q'$-invariance, that is manifest in the present lattice
formulation, possibilities such as $\tr X_\mu$ are excluded. Then only possible
combinations are, 1~(identity), $\tr\mathcal{F}_{01}$ and~$\tr D'$.
The identity operator~1 has no dynamical effect, while
$\int d^2x\,\tr\mathcal{F}_{01}=\int d^2x\,\tr F_{01}$ and
$\int d^2x\,\tr D'$ are simply the theta term and the FI term, respectively.

Finally, terms being proportional to the third term in the parentheses arise at
the two-loop level or lower. A unique local operator with~$p\leq0$ is the
identity~1 and thus this has no dynamical effect.

In this way, we observe that only nontrivial local terms that can radiatively
be generated in the effective action are the FI term and the theta term. These
are terms already present in the continuum classical action and of course
invariant under all symmetries of the target continuum theory, especially under
supersymmetry. Therefore, there is no need to supplement local counterterms to
restore symmetries of the continuum theory; this shows that in perturbation
theory symmetries are restored in the continuum limit without any fine tuning.

With the present lattice regularization, the radiative correction to the FI
parameter~$\kappa$ up to the one-loop order is given by
\begin{equation}
   \kappa_R=\kappa+\frac{1}{4\pi}
   \left[N\left(\ln a^2-\ln32\right)
   +\sum_{I=1}^N\ln(\tilde m_{+I}^*\tilde m_{+I})\right],
\label{threexthirtyfive}
\end{equation}
for $a\to0$. Higher order corrections are UV finite and of~$O(g^2/m^2)$, where
$m^2$ is a linear combination of $\tilde m_{+I}^*\tilde m_{+I}$. Thus, in order
to take the continuum limit while keeping $\kappa_R$ fixed, one has to set
$\kappa=(N/4\pi)\ln\beta+\text{const}$.\footnote{
A possible renormalization scheme for $\kappa$ is given by setting
$\tilde m_{+I}^*\tilde m_{+I}=2\mu^2$ in one-loop
expression~(\ref{threexthirtyfive}), where $\mu$~is a renormalization scale.
If one defines the lambda parameter in this scheme (this scheme corresponds to
the choice in~Ref.~\cite{Witten:1993xi}) by
$\Lambda^{\text{MASS}}\equiv\mu e^{-(2\pi/N)\kappa_R(\mu)}$, then the ratio to
the lambda parameter on the lattice
$\Lambda^{\text{LAT}}\equiv(1/a)e^{-(2\pi/N)\kappa}$ is given by
$\Lambda^{\text{MASS}}/\Lambda^{\text{LAT}}=4$ for any $N\neq0$,
while $\Lambda^{\overline{\text{MS}}}/\Lambda^{\text{MASS}}=\sqrt{2}$.}%

On the other hand, according to a standard argument on the basis of the
$U(1)_A$ anomaly, the theta parameter would radiatively be corrected as
\begin{equation}
   \vartheta_R=\vartheta
   +\frac{1}{2i}\sum_{I=1}^N
   \ln\left(\frac{\tilde m_{+I}}{\tilde m_{+I}^*}\right),
\end{equation}
for $a\to0$. Since this~$\vartheta_R$ is UV finite, $\vartheta$ can be taken to
be independent of the lattice spacing in the continuum limit.

\section{Conclusion}
In this paper, we presented a lattice formulation of 2D $\mathcal{N}=(2,2)$
$U(k)$ SQCD with $N$ fundamental matter multiplets. The formulation uses
compact gauge link variables and respects one exact fermionic symmetry~$Q'$
on the 2D regular lattice. Our lattice action is considerably simpler compared
with the lattice action of~Ref.~\cite{Kikukawa:2008xw}. In particular, the
lattice action is polynomial in bosonic variables $U_\mu(x)$, $V_\mu(x)$ and
their inverse matrices. We think that this point can be a practical advantage
in actual implementations on the computer. In the near future, we hope to use
the present lattice formulation to investigate physical questions (as
in~Refs.~\cite{Kanamori:2007ye,Kanamori:2007yx,Kanamori:2008vi,Kanamori:2008yy,
Kanamori:2009dk}) in 2D $\mathcal{N}=(2,2)$ $U(k)$ SQCD with $N$ fundamental
multiplets.%

Still, at the present moment, the range of applicability of the present
formulation is rather limited compared with the formulation
of~Ref.~\cite{Kikukawa:2008xw}; we could not incorporate the superpotential and
we do not know how to truncate the gauge group $U(k)$ to~$SU(k)$. Also,
although it is almost straightforward to define a nilpotent lattice
$Q'$-transformation and a $Q'$-exact lattice action for matter multiplets in
other gauge representations (such as the anti-fundamental and the adjoint), we
could not find a rigorous argument that shows that $U_\mu(x)=1$ (up to gauge
transformations) is a unique weak coupling saddle point. Further study is
needed on these possible generalizations of the present lattice formulation.

D.~K.\ and H.~S.\ would like to thank Issaku Kanamori for discussions.
H.~S.\ would like to thank Martin L\"uscher for a helpful remark.
This work was initiated when two of us (F.~S.\ and~H.~S.) attended the Niels
Bohr International Academy workshop, ``Lattice Supersymmetry and Beyond''. We
would like to thank participants, especially, Simon Catterall, Alessandro
D'Adda, Michael Endres, Noboru Kawamoto, So Matsuura for useful information and
Poul Henrik Damgaard and Hidenori Fukaya for the hospitality extended to us at
the Niels Bohr Institute.
The work of H.~S.\ is supported in part by a Grant-in-Aid for Scientific
Research, 18540305.



\appendix
\section{Flat connections on a periodic lattice}
\label{app:a}
In this appendix, we give the most general solution of $U_{01}(U)(x)=1$ on a
periodic lattice.

Let $W_\mu$ be the products of gauge link variables along nontrivial cycles
on a periodic lattice (Wilson lines):
\begin{align}
   W_0&\equiv U_0(0,0)U_0(a,0)\dots U_0((N_0-1)a,0),
\nonumber\\
   W_1&\equiv U_1(0,0)U_1(0,a)\dots U_1(0,(N_1-1)a).
\end{align}
Since $[W_0,W_1]=0$ from $U_{01}(x)=1$, there exists $\Omega\in U(k)$ such that
\begin{equation}
   \Omega^{-1}W_0\Omega=
   \begin{pmatrix}
   e^{i\alpha_1}& &\\
   &\ddots&\\
   & &e^{i\alpha_k}
   \end{pmatrix},\qquad
   \Omega^{-1}W_1\Omega=
   \begin{pmatrix}
   e^{i\beta_1}& &\\
   &\ddots&\\
   & &e^{i\beta_k}
   \end{pmatrix},
\label{axtwo}
\end{equation}
where $0\leq\alpha_i,\beta_i<2\pi$ ($i=1$, \dots, $k$).
Set the gauge transformation function at the origin
\begin{equation}
   g(0,0)=\Omega,
\end{equation}
and let $T_\mu$ be the $N_\mu$-th root of the right-hand side
of~Eq.~(\ref{axtwo}),
\begin{equation}
   T_0\equiv
   \begin{pmatrix}
   e^{i\alpha_1/N_0}& &\\
   &\ddots&\\
   & &e^{i\alpha_k/N_0}
   \end{pmatrix},\qquad
   T_1\equiv
   \begin{pmatrix}
   e^{i\beta_1/N_1}& &\\
   &\ddots&\\
   & &e^{i\beta_k/N_1}
   \end{pmatrix}.
\label{axfour}
\end{equation}
Clearly, $[T_0,T_1]=0$. Finally, we define the gauge transformation function
$g(x)$ by
\begin{equation}
   g(x)^{-1}\equiv T_0^{-C_0}T_1^{-C_1}g(0,0)^{-1}U(C),
\end{equation}
where $C$ denotes a certain path on the periodic lattice that connects the
origin~$(0,0)$ and the point~$x$, and $U(C)$ is the path ordered product of link
variables along~$C$. $C_\mu$~are integers defined by
\begin{equation}
   C_\mu\equiv(\text{$\sharp$ of $U_\mu$ in $U(C)$})
   -(\text{$\sharp$ of $U_\mu^{-1}$ in $U(C)$}).
\end{equation}
Because of $U_{01}(x)=1$, $g(x)$ defined above does not depend on the chosen
path from~$(0,0)$ to~$x$. It is then straightforward to confirm that $g(x)$
is periodic on the lattice and
\begin{equation}
   g(x)^{-1}U_\mu(x)g(x+a\hat\mu)=T_\mu,
\end{equation}
which shows that the most general solution of $U_{01}(U)(x)=1$ is given by
$U_\mu(x)=g(x)T_\mu g(x+a\hat\mu)^{-1}$.

\end{document}